\documentclass[pra,aps,superscriptaddress, twocolumn]{revtex4}
\usepackage{bm,dcolumn,amsmath,graphicx,amsfonts,amssymb,lipsum}
\usepackage{epsfig}
\usepackage[dvipsnames]{xcolor}
\usepackage{CJKutf8}

\newcommand{\jgli}[1]{{\color{blue}{\bf #1}}}
\begin{document}
\begin{CJK*}{UTF8}{gkai}
	
\title{Theoretical study of electronic structure of erbium and fermium}
% ($Z=100$): energy levels, transition rates and isotope shift}
%Calculation of energy levels and electric dipole transition rates for Er and Fm}
	\author{Saleh O. Allehabi}
      \affiliation{School of Physics, University of New South Wales, Sydney 2052, Australia}
	\author{Jiguang Li (李冀光)}
%	\author{Jiguang Li (???)}
	\affiliation{Institute of Applied Physics and Computational Mathematics, Beijing 100088, China}
	\author{V. A. Dzuba}
	\affiliation{School of Physics, University of New South Wales, Sydney 2052, Australia}
	\author{V. V. Flambaum}
	\affiliation{School of Physics, University of New South Wales, Sydney 2052, Australia}
	\affiliation{Helmholtz Institute Mainz, Johannes Gutenberg University, 55099 Mainz, Germany}
		
	\date{\today}
	
\begin{abstract}
We use a recently developed version of the configuration method for open shells to study electronic structure of erbium and fermium atoms. We calculate excitation energies of odd states connected to the even ground state by electric dipole transitions, the corresponding transition rates, isotope shift, hyperfine structure, ionization potentials and static scalar polarizabilities. We argue that measuring isotope shift for several transitions can be used to study nuclear deformation in even-even nuclei. This is important for testing nuclear theory and for searching for the hypothetical island of stability. Since erbium and fermium have similar electronic structures, calculations for erbium serve as a guide to the accuracy of the calculations.
% We use similar calculations for erbium, which lighter analog of fermium, to validate the method and estimate the accuracy of the calculations. 
\end{abstract}
	
\maketitle
\end{CJK*}

\section{Introduction}

Fermium is one of the two heaviest atoms for which experimental spectroscopic data are available. It was intensively studied in the last two decades both theoretically and experimentally \cite{Liu1,Liu2,Cao,Moritz,Fm1,Fm2,Wendt,Sato}.
Frequencies of seven electric dipole transitions from the ground state to excited odd states have been measured~\cite{Fm1,Fm2}, and the first ionization potential has been determined~\cite{Wendt,Sato}. The measurements were led by theoretical predictions  made with the use of the multi-configurational Dirac-Fock method (MCDF)~\cite{Fm1}. It is clear from comparing these data to the spectrum of the erbium atom, which is a lighter analog of Fm, that a large number of Fm states still escaped experimental detection and theoretical determination. Further study of Fm is needed to understand its electronic structure and its difference from Er caused by interplay between correlation and relativistic effects.

The only other atom, heavier than Fm for which spectroscopic data are available is nobelium. Frequency of just one electric dipole transition between the ground state singlet $7s^2 \ ^1$S$_0$ and the excited odd-parity singlet $7s7p \ ^1$P$^{\rm o}_1$ has been measured for three isotopes, $^{252,253,254}$No~\cite{No-Nature,No-IS}. These measurements led also to determinations of the isotope shifts (IS) between these isotopes and the hyperfine structure (hfs) of the $^{253}$No isotope. The study of hfs combined with atomic calculations~\cite{No-hfs} allows extraction of the values of nuclear magnetic dipole and electric quadrupole moments. The large value of the electric quadrupole moment indicates a strong nuclear deformation in the $^{253}$No isotope. The information about nuclear deformation is important in search of the island of stability: metastable nuclei are expected to be spherically symmetric. The use of hfs to study nuclear deformation works only for odd isotopes. Even-even isotopes as a rule have zero nuclear spin, and thus no hfs of atomic levels.
It is worth mentioning that nuclear theory is more complicated and less accurate for odd isotopes than for even-even isotopes~\cite{Afanasjev}.    
	
In our previous works we suggested to use IS measurements to study nuclear deformation in even-even isotopes~\cite{E120,No-nuc}. Similar to how the change of nuclear root-mean square (RMS) radius ($\delta \langle r^2 \rangle$) can be obtained from the measurements of IS in an atomic transition, the change of both, $\langle r^2 \rangle$ and the parameter of nuclear quadrupole deformation $\beta$ can be found from IS measurements made for at least two atomic transitions. This approach can be used for fermium as well. There are 20 known Fm isotopes from $^{241}$Fm to $^{260}$Fm~\cite{NUBASE2016}. All even-even isotopes in this range are strongly deformed~\cite{Afanasjev}, and the parameter $\beta$ varies significantly from isotope to isotope. The maximum change is between $^{260}$Fm ($\beta=0.265$) and $^{248}$Fm or $^{250}$Fm ($\beta=0.296$), i.e. $\Delta \beta_{\rm max}=0.031$. Such a large change is relatively easy to detect. In this work we calculate parameters of the sensitivity of IS to the change in nuclear parameters,  $\langle r^2 \rangle$ and $\beta$.
	
The ground state of Fm is $5f^{12}7s^2 \ ^3$H$_6$. The seven odd states for which the measurements were done all belong to the same configuration $5f^{12}7s7p$~\cite{Fm1,Fm2}. However, it is well known that erbium, which is a lighter analog of Fm, has a very rich spectrum, with most of the known odd states belonging to two configurations, $4f^{12}6s6p$ and $4f^{11}6s^25d$~\cite{NIST}. It is natural to expect similar features in Fm. In this work we perform similar calculations for both atoms. Calculations for Er mostly serve as a guide to the accuracy of calculations. Calculations for Fm demonstrate that both atoms have many properties in common. Some differences in the spectra can be explained by stronger relativistic effects in Fm. We calculate energy separations for the odd excited states connected to the ground state by electric dipole transitions, and the corresponding transition rates to identify stronger transitions easiest to detect. Our calculations are compared with the measurements and previous MCDF calculations. 
	
\section{Method of calculation}

We use the recently developed configuration interaction with perturbation theory (CIPT) method~\cite{cipt} to perform the calculations. The method was developed specifically for atoms with open shells that have a large number of electrons in open shells. Correlations between these electrons are treated using the configuration interaction technique. A set of single-electron basis states used for these calculations forms the valence space.  The CIPT method was successfully used for atoms with open $f$~\cite{Yb, Md}, $d$~\cite{6d} and $p$~\cite{e118} shells. The maximum number of electrons in the valence space was sixteen (Yb and No~\cite{No-IS,No-hfs,Yb}). The main idea of the method is neglecting off-diagonal matrix elements between high-energy states in the CI matrix. The idea is used in several similar approaches~\cite{MCDFoff,Berengut,Kahl}. However, in the CIPT method one more step is made, the whole CI matrix is reduced to the effective CI matrix of a much smaller size, in which matrix elements between low-lying states \textit{i} and \textit{j} are corrected  by the expression similar to the second-order perturbative correction to the energy:
\begin{equation}
\langle i|H^{CI}|j\rangle\rightarrow\langle i|H^{CI}|j\rangle+\sum_{k}\frac{\langle i|H^{CI}|k\rangle\langle k|H^{CI}|j\rangle}{{E}-E_{k}} \jgli{,}
\label{e:CIPT}
\end{equation}
where $E$ refers to  the energy of the state of interest, and $E_k$ denotes the diagonal matrix element for high-energy states, $E_k =\langle k| H^{CI}| k\rangle$. Summation in (\ref{e:CIPT}) goes over all high-energy states. 
The ordering of the states on the energy scale is done by the values of $E_k$. This ordering can be justified by moving more states from the perturbative term (second term in (\ref{e:CIPT})) to the effective CI matrix. The energies $E$ and wave functions $X$ are found by solving the matrix eigenvalue problem
\begin{equation}
\left(H^{CI}-E{I}\right) X = 0,
\label{e:CI} 
\end{equation}
with $H^{CI}$ matrix given by (\ref{e:CIPT}); where $I$ is the unit matrix. Reducing the matrix size by (\ref{e:CIPT}) does not affect the resulting energies as long as the energy $E$ is the same in (\ref{e:CIPT}) and (\ref{e:CI}). Since the energy is not known in advance, iterations over energy are needed. Usually five to ten iterations are enough for full convergence.

We use the B-spline technique~\cite{B-spline} to build a single-electron basis set. These states are constructed as linear combinations of B-splines that are eigenstates of the relativistic Hartree-Fock (HF) Hamiltonian with the $V^{N-1}$ potential. The self-consistent HF procedure is first done for an atom with one electron removed. For instance, the ground state of Er belongs to the [Xe]$4f^{12}6s^2$ configuration. The HF procedure is done for the [Xe]$4f^{12}6s$ configuration, and basis states for valence electrons are calculated in the resulting $V^{N-1}$ potential. Similarly, the [Rn]$5f^{12}7s$ configuration is used in the HF calculations for Fm. We use forty B-splines confined to a sphere of the radius $R_{\rm max} = 40~a_B$. Fourteen out of forty lowest-energy eigenstates of the HF Hamiltonian are used in the CIPT calculations. Higher-lying states give only negligible contributions due to large energy denominators in (\ref{e:CIPT}). Many-electron basis states for the CI calculations are constructed by exciting one or two electrons from initial reference valence configurations. For example, to calculate even states of Er, we use the $4f^{12}6s^2$ configuration as a reference. All states of this configuration are used to construct the effective CI matrix. All states obtained by single and double excitations are used in the perturbative term, which is the last term in (\ref{e:CIPT}). For odd states, we use four reference configurations $4f^{12}6s6p$, $4f^{12}6s7p$, $4f^{11}6s^25d$ and $4f^{12}6s5f$. All states from these three configurations go to the effective CI matrix while all states obtained by exciting electrons from these configurations go to the perturbative term. Similarly for Fm, the reference configurations are $5f^{12}7s^2$ for even states and $5f^{12}7s7p$, $5f^{12}7s8p$, $5f^{11}7s^26d$ and $5f^{12}7s6f$ for odd states. 

The calculations are fully relativistic. Our single-electron operator in the HF and CI Hamiltonians comes from the Dirac equation. Moreover, the Breit and quantum electrodynamic corrections are included similar to what was done in our previous works~\cite{Yb,6d,Md,e118}.

To calculate amplitudes of electric dipole transitions and hyperfine structure, we need to include an external field in the equations. This is the electric field of a photon in case of electric dipole transitions or nuclear magnetic or quadrupole electric field in case of hfs.
We use the time-dependent Hartree-Fock method~\cite{CPM} (equivalent to the random-phase approximation (RPA)) to do this. The RPA equations are first solved for the atom in the same $V^{N-1}$ approximation as in the HF calculations. The RPA equations
\begin{equation}
\left( H^{HF} -\epsilon_i\right)\delta \psi_i = -\left(\hat d +\delta V^{N-1}\right)\psi_i
\label{e:RPA}
\end{equation}
are iterated for all atomic states $i$ in the core to find the correction to the atomic potential $\delta V^{N-1}$ caused by the effect of an external field.
Transition amplitudes are calculated as
\begin{equation}
A_{ab} = \langle a | \hat d + \delta V^{N-1} | b \rangle,
\label{e:A}
\end{equation}
while hfs constants are found from the diagonal matrix elements.
Here $|a\rangle$ and $|b\rangle$ are many-electron states obtained in the CI calculations (\ref{e:CI}), $\hat d$ is the electric dipole operator (we use the length form, $\hat d = -e\sum_n  r_n$) in case of electric dipole transitions or operator of nuclear magnetic dipole or electric quadrupole field in case of hfs. The rate of spontaneous emission of a photon in the transition from state $b$ to state $a$ is given by (in atomic units)
\begin{equation}
T_{ab} = \frac{4}{3}\left(\alpha\omega_{ab}\right)^3\frac{A^{2}_{ab}}{2J_b+1},
\label{e:T}
\end{equation}
where $\alpha$ is the fine structure constant, $\omega_{ab}$ is the frequency of the transition.

The results for energy levels and transition rates for Er are presented in Table~\ref{t:Er} and compared to experiment. We see that the difference between theory and experiment for the energies is about a few hundred cm$^{-1}$ for the states of the $4f^{12}6s6p$ configuration and up to 3000~cm$^{-1}$ for $4f^{11}6s^25d$. Similar accuracy can be expected for Fm. Note that the difference in the spectra of Er and Fm mostly comes from relativistic effects, while the main source of numerical uncertainty is the incomplete treatment of correlations. Since Er and Fm have similar electronic structure, it is natural to assume that the correlations in both atoms are also similar. The same assumption worked well in many earlier calculations (see, e.g.~\cite{Md,e118,6d}).

A comparison of transition rates with available experimental data (see Table~\ref{t:Er}) shows that the accuracy of calculations is significantly lower than for the energies. 
Most probably this is due to the fact that the high-energy states (those participating in the last term in (\ref{e:CIPT})) are not included into the resulting wave function in the current version of the computer code. In other words, perturbation theory correction is considered to correct the energies but not wave functions. Wave functions are still constructed from the reference configurations only (those included in the first part of the effective CI Hamiltonian (\ref{e:CIPT})). In principle, it is possible to correct the wave function, too. This might be a subject for future work.

\begin{table*}
		
		\caption{\label{t:Er}
			Excitation energies ($E$, cm$^{-1}$), transition amplitudes ($A$$_{ab}$, a.u.), electric dipole transition rates to the ground state ($T$$_{ab}$, s$^{-1}$), and $g$-factors for some low odd states of Er atom with \textit{J}= 5,6,7. The discrepancy between the NIST and present energies is given by $\Delta$ = $\textit E$$_{\rm NIST}$ - ${\textit E}$$_{\rm present}$.} 
		\begin{ruledtabular}
			\begin{tabular}{cccccccccccc}
				&&&&
				\multicolumn{3}{c}{$E$}&
				\multicolumn{1}{c}{$A$$_{ab}$}&
				\multicolumn{2}{c}{$T$$_{ab}$}&
				\multicolumn{2}{c}{\textit{g}}\\
				\cline{5-7}
				\cline{8-8}
				\cline{9-10}
				\cline{11-12}
				\multicolumn{1}{c}{\textit{N}}& 
				\multicolumn{1}{c}{Conf.}&
				\multicolumn{1}{c}{Term}&
				\multicolumn{1}{c}{\textit{J}}&
				\multicolumn{1}{c}{Present}&
				\multicolumn{1}{c}{NIST \tablenotemark[1]}&
				\multicolumn{1}{c}{ $\Delta$ }&
				\multicolumn{1}{c}{Present}&
				\multicolumn{1}{c}{{Present}}&
				\multicolumn{1}{c}{Expt. \tablenotemark[2]}&
				\multicolumn{1}{c}{Present}&
				\multicolumn{1}{c}{NIST \tablenotemark[1]}\\
				\hline
				1 & $4f^{12}6s^2$    & $^3$H$$ &     6 &0&0&0&0&0&0& 1.1651  & 1.16381 \\
				2 &4f$^{11}$($^4$I$^o$$_{15/2}$)5d$_{3/2}$6s$^2$& $(15/2, 3/2)^o $ & 6 & 4719 & 7177  &2458&0.0012& 2.24$\cdot$10$^{-02}$& & 1.3067& 1.302 \\
				3 &4f$^{11}$($^4$I$^o$$_{15/2}$)5d$_{3/2}$6s$^2$& $(15/2, 3/2)^o $ & 7 & 5455 & 7697 &2242 &0.0034& 2.54$\cdot$10$^{-01}$& & 1.2623& 1.266 \\
				4 &4f$^{11}$($^4$I$^o$$_{15/2}$)5d$_{5/2}$6s$^2$& $(15/2, 5/2)^o $ & 7 & 8072 & 11888 &3816 &0.2788& 5.52$\cdot$10$^{+03}$& & 1.1477& 1.153 \\
				5 &4f$^{11}$($^4$I$^o$$_{15/2}$)5d$_{5/2}$6s$^2$& $(15/2, 5/2)^o $ & 5 & 8129 & 11401&3272  &-0.1370& 1.85$\cdot$10$^{+03}$& & 1.2030& 1.205 \\
				6 &4f$^{11}$($^4$I$^o$$_{15/2}$)5d$_{5/2}$6s$^2$& $(15/2, 5/2)^o $ & 6 & 8572 & 11800&3228  &0.3254& 1.04$\cdot$10$^{+04}$& & 1.1779& 1.190 \\
				7 &4f$^{11}$($^4$I$^o$$_{13/2}$)5d$_{3/2}$6s$^2$& $(13/2, 3/2)^o $ & 5 & 13476 & 15185&1709  &-0.3621& 5.91$\cdot$10$^{+04}$& & 1.1914& 1.160 \\
				8 &4f$^{11}$($^4$I$^o$$_{13/2}$)5d$_{3/2}$6s$^2$& $(13/2, 3/2)^o $ & 7 & 14214 & 15847&1633  &-0.5442& 1.15$\cdot$10$^{+05}$& & 1.0686& 1.070 \\
				9 &4f$^{11}$($^4$I$^o$$_{13/2}$)5d$_{3/2}$6s$^2$& $(13/2, 3/2)^o $ & 6 & 14265&16070&1805 &0.9524& 4.10$\cdot$10$^{+05}$&(9.2$\pm$0.05)$\cdot$10$^{+05}$& 1.1497& 1.200 \\
				10 &4f$^{11}$($^4$I$^o$$_{13/2}$)5d$_{5/2}$6s$^2$& $(13/2, 5/2)^o $ & 7 & 15233 & 17796&2563  &0.4817& 1.11$\cdot$10$^{+05}$& & 1.1184& 1.110 \\
				11 &4f$^{11}$($^4$I$^o$$_{13/2}$)5d$_{5/2}$6s$^2$& $(13/2, 5/2)^o $ & 6 & 15278 & 17456&2178  &-0.1008& 5.64$\cdot$10$^{+03}$& & 1.0650& 1.070 \\
				12 &4f$^{11}$($^4$I$^o$$_{13/2}$)5d$_{5/2}$6s$^2$& $(13/2, 5/2)^o $ & 5 & 15335 & 17029 &1694 &0.0355& 8.34$\cdot$10$^{+02}$& & 1.1042& 1.150 \\
				13 &$4f^{12}(^3$H$_6)6s6p(^3$P$^o_0)$& $(6,0)^o $ & 6 & 16499 & 16321 &-178 &-0.5998& 2.52$\cdot$10$^{+05}$& & 1.2860& 1.220 \\
				14 &4f$^{11}$($^4$I$^o$$_{11/2}$)5d$_{3/2}$6s$^2$& $(11/2, 3/2)^o $ & 7 & 16697 & 18774&2077 &-0.2504 & 3.94$\cdot$10$^{+04}$& & 0.9480& 0.965 \\
				15 &$4f^{12}(^3$H$_6)6s6p(^3$P$^o_1)$& $(6,1)^o $& 6 & 17131 & 17074  &-57&1.1178& 9.78$\cdot$10$^{+05}$& & 1.0369& 1.070 \\
				16 &$4f^{12}(^3$H$_6)6s6p(^3$P$^o_1)$& $(6,1)^o $& 7 & 17316 & 17157  &-159&1.6462& 1.90$\cdot$10$^{+06}$&(1.17$\pm$0.06)$\cdot$10$^{+06}$ & 1.1879& 1.195 \\
				17 &4f$^{11}$($^4$I$^o$$_{11/2}$)5d$_{3/2}$6s$^2$& $(11/2, 3/2)^o $ & 6 & 17383 & 19508&2125 &0.0816 & 5.45$\cdot$10$^{+03}$& & 0.9874& 0.960 \\
				18 &$4f^{12}(^3$H$_6)6s6p(^3$P$^o_1)$& $(6,1)^o $ & 5 & 17425 & 17348  &-77&1.3160& 1.69$\cdot$10$^{+06}$&(8.4$\pm$0.04)$\cdot$10$^{+05}$ & 1.1685& 1.175 \\
				19 &4f$^{11}$($^4$I$^o$$_{11/2}$)5d$_{3/2}$6s$^2$& $(11/2, 3/2)^o $ & 5 & 18021 & 19563&1542 &0.1200 & 1.55$\cdot$10$^{+04}$& & 0.9687& 0.990 \\
				20 &4f$^{11}$($^4$I$^o$$_{11/2}$)5d$_{5/2}$6s$^2$& $(11/2, 5/2)^o $ & 7 & 18962 & 21168 &2206 &0.6728& 4.16$\cdot$10$^{+05}$& & 1.0557& 1.065 \\
				21&$4f^{12}(^3$H$_6)6s6p(^3$P$^o_2)$& $(6,2)^o $& 7& 18987 & 19125  &138&0.0613& 3.47$\cdot$10$^{+03}$& & 1.2448& 1.235 \\
				22 &$4f^{12}(^3$H$_6)6s6p(^3$P$^o_2)$& $(6,2)^o $& 6 & 19164 & 19327  &163&-1.0401& 1.19$\cdot$10$^{+06}$& & 1.0408& 1.180 \\
				23 &$4f^{12}(^3$H$_6)6s6p(^3$P$^o_2)$& $(6,2)^o $ & 5 & 19203 & 19201 &-2 &-0.6045& 4.76$\cdot$10$^{+05}$& & 1.0464& 1.060 \\
				24 &4f$^{11}$($^4$I$^o$)5d6s$^2$& $^o $ & 6 & 19304 & 20738  & 1434&-0.3159&1.12$\cdot$10$^{+05}$& & 0.9509& 0.855 \\
				25 &4f$^{11}$($^4$I$^o$$_{11/2}$)5d$_{5/2}$6s$^2$& $(11/2, 5/2)^o $ & 5& 19435 & 21393&1958 &-0.5741 & 4.45$\cdot$10$^{+05}$& & 0.9631& 1.005 \\
				26 &4f$^{11}$($^4$I$^o$$_{11/2}$)5d$_{5/2}$6s$^2$& $(11/2, 5/2)^o $& 6&19628&21702&2074&-1.2573& 1.86$\cdot$10$^{+06}$&(7.1$\pm$0.4)$\cdot$10$^{+06}$ & 1.0318& 1.055 \\
				27 &4f$^{11}$($^4$I$^o$)5d6s$^2$& $^o $ & 5 & 19926 & 20917  & 991&0.4380&2.79$\cdot$10$^{+05}$& & 0.9062& 0.980 \\
				28 &4f$^{11}$($^4$F$^o$)5d6s$^2$& $^o $ & 6 & 21348 & 22584  & 1236&1.5116&3.46$\cdot$10$^{+06}$&(2.55$\pm$0.13)$\cdot$10$^{+06}$ & 1.0072& 1.130 \\
				29 &4f$^{11}$($^4$I$^o$)5d6s$^2$& $^o $ & 5 & 21843 & 22673  & 830&1.9920&7.61$\cdot$10$^{+06}$&(5.52$\pm$0.28)$\cdot$10$^{+06}$ & 1.0235& 1.040 \\
				30 &4f$^{11}$($^4$I$^o$)5d6s$^2$& $^o $ & 7 & 22566 & 23081  & 515&0.7092&7.80$\cdot$10$^{+05}$& & 1.0575& 1.010 \\
				31 &$4f^{12}(^3$F$)6s6p(^3$P$^o)$& $^o $ & 5 & 23237 & 22124  & -1113&-0.2492&1.43$\cdot$10$^{+05}$& & 1.1832& 1.285 \\
				32 &$4f^{12}(^3$H$)6s6p$& $^o $ & 5 & 23422 & 23447  & 25&-0.5999&8.51$\cdot$10$^{+05}$& & 1.0940& 1.080 \\
				33 &$4f^{12}(^3$F$)6s6p(^3$P$^o)$& $^o $ & 5 & 23801 & 23856  & 55&3.0144&2.25$\cdot$10$^{+07}$&(6.6$\pm$0.3)$\cdot$10$^{+06}$ & 1.0578& 1.140 \\
				34 &$4f^{12}(^3$H$)6s6p$& $^o $ & 5 & 23872 & 23885  & 13&4.7735&5.70$\cdot$10$^{+07}$& & 1.0665& 1.100 \\
				35 &$4f^{12}(^3$F$)6s6p(^3$P$^o)$& $^o $ & 6 & 24107 & 23831  & -276&-1.6371&5.85$\cdot$10$^{+06}$& & 1.0939& 1.250 \\
				36 &4f$^{11}$($^4$I$^o$)5d6s$^2$& $^o $ & 6 & 24393 & 24457  & 64&-3.7556&3.19$\cdot$10$^{+07}$&(3.26$\pm$1.6)$\cdot$10$^{+07}$ & 1.0911& 1.050 \\
				37 &$4f^{12}(^3$H$)6s6p$& $^o $ & 5 & 24587 & 24083  & -504&-5.1389&7.22$\cdot$10$^{+07}$&(10.2$\pm$5.)$\cdot$10$^{+07}$ & 1.1254& 1.128 \\
				38 &$4f^{12}(^3$H$)6s6p(^3$P$^o)$& $^o $ & 6 & 24674 & 24246  & -428&-0.2660&1.66$\cdot$10$^{+05}$& & 1.2273& 1.085 \\
				39 && $^o $ & 5 & 24791 & 25364 &573&-3.6033 & 3.64$\cdot$10$^{+07}$& & 1.1948& 1.180 \\
				40 &4f$^{11}$($^4$F$^o$)5d6s$^2$& $^o $ & 6 & 24816 & 25393  &577& -7.0787&1.19$\cdot$10$^{+08}$&(3.19$\pm$1.6)$\cdot$10$^{+07}$& 1.1267& 1.075 \\
				41 &$4f^{12}(^3$H$)6s6p$& $^o $ & 7 & 24845 & 24943  & 98&10.6390&2.34$\cdot$10$^{+08}$&(1.85$\pm$10.)$\cdot$10$^{+08}$ & 1.1458& 1.160 \\
				42 && $^o $ & 7 & 25379 & 25159  & -220&-4.7652&5.01$\cdot$10$^{+07}$&(4.03$\pm$2.1)$\cdot$10$^{+07}$ & 1.1613& 1.170 \\
				43 &$4f^{12}(^3$H$)6s6p$& $^o $ & 6 & 25783 & 25880  & 97&-2.0255&1.09$\cdot$10$^{+07}$&(1.22$\pm$6.)$\cdot$10$^{+08}$ & 1.1534& 1.150 \\
				44 &$4f^{12}(^3$H$)6s6p(^3$P$^o)$& $^o $ & 5 & 25888 & 26199  & 311&-0.1108&3.92$\cdot$10$^{+04}$& & 1.0683& 1.045 \\
				45 &4f$^{11}$($^4$I$^o$)5d6s$^2$& $^o $ & 5 & 26202 & 25163  & -1039&4.1159&5.61$\cdot$10$^{+07}$&(3.76$\pm$1.9)$\cdot$10$^{+07}$ & 1.0100& 1.175 \\
				46 && $^o $ & 7 & 26874 & 27231&357 &-0.4816 & 6.07$\cdot$10$^{+05}$&& 1.1318& 1.135 \\
				
\end{tabular}
%			\begin{minipage}{6.5 cm}%
%				\tiny  $^{a}$ Ref.~\cite{NIST}; $^{b}$Ref.~\cite{Martin}.%
%			\end{minipage}%
\end{ruledtabular}
\tablenotetext[1]{Ref.~\cite{NIST}}

\tablenotetext[2]{Ref.~\cite{Lawler}}

\end{table*}

Table~\ref{t:Er} also presents the values of calculated and experimental Land\'{e} $g$-factors. The $g$-factors are useful for identification of states. In some cases %(e.g., when no experimental values are available) 
it is useful to compare the calculated $g$-factors to a non-relativistic expression
\begin{equation}
g_{NR} = 1 + \frac{J(J+1)-L(L+1)+S(S+1)}{2J(J+1)}.
\label{e:g}
\end{equation}
This expression is convenient to use in a non-relativistic limit when total orbital momentum $L$ and total spin $S$ are good quantum numbers ($J$ is total angular momentum, $\mathbf{J} =\mathbf{L}+\mathbf{S}$). This is not the case for both Er and Fm. Note however, that non-relativistic labelling of the states can be used at least in principle in a relativistic case, too. It can be done by expanding a relativistic wave function over a non-relativistic basis and taking  the values of $L$ and $S$ that correspond to the largest term in the expansion. This was done in previous calculations for Fm~\cite{Fm1}. Using their notations and formula (\ref{e:g}) and comparing the obtained values with our calculated $g$-factors helps to link the results of our calculations to those of Ref.~\cite{Fm1}.

\section{Results}

\begin{figure}[tb]
\epsfig{figure=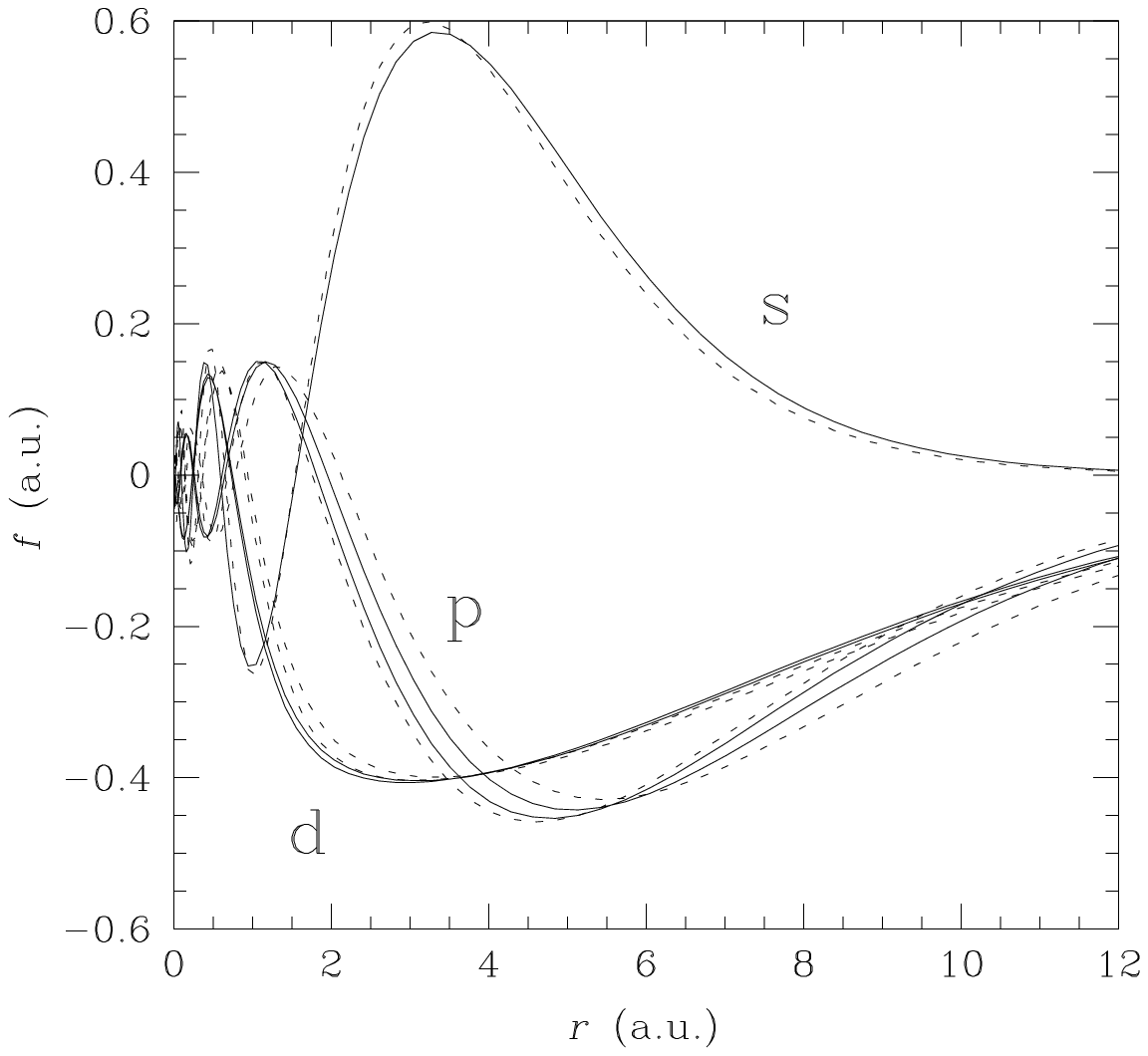,scale=0.4}
\caption{{Upper components $f(r)$ (see Eq.~(\ref{e:orb})) of the lowest valence} $6s_{1/2}, 6p_{1/2}, 6p_{3/2}, 5d_{3/2}, 5d_{5/2}$ orbitals of Er (solid lines) and $7s_{1/2}, 7p_{1/2}, 7p_{3/2}, 6d_{3/2}, 6d_{5/2}$ orbitals of Fm (dot lines).}
\label{f:ErFm}
\end{figure}

The results for Fm are presented in Table~\ref{t:Fm}. As expected, the spectrum of Fm is very similar to Er. There are some differences, mostly caused by relativistic effects, which are expected to be about two times larger in Fm than in Er. To understand the difference, we compare the outermost single-electron $7s,7p$ and $6d$ orbitals of Fm to the $6s,6p$ and $5d$ orbitals of Er. The orbitals have a form
\begin{equation}\label{e:orb}
\psi (\mathbf{r}) = \frac{1}{r}\left(\begin{array}{c} f(r) \Omega_{jlm}(\mathbf{n}) \\ i\alpha g(r)\Omega_{j \tilde{l}m}(\mathbf{n}) \end{array} \right).
\end{equation}
Here $f(r)$ and $g(r)$ are the upper and lower radial components of the wave function, $\mathbf{n}=\mathbf{r}/r$, $\alpha$ is the fine structure constant, $\tilde{l}= l-2j$, $\Omega_{jlm}(\mathbf{n})$ is a spherical spinor.
Fig.~\ref{f:ErFm} shows the upper components $f(r)$ of the Er and Fm orbitals. Stronger relativistic effects of Fm move the $7s$ and $7p_{1/2}$ orbitals closer to the nucleus than similar $6s$ and $6p_{1/2}$ orbitals in Er. This is because relativistic effects are stronger at short distances from the nucleus, where densities of the $s$ and $p_{1/2}$ states are not negligible. On these distances relativistic effects associated with the Dirac equation act as an attraction to the nucleus. In contrast, the densities of the  $p_{3/2}$, $d_{3/2}$ and $d_{5/2}$ states are small near the nucleus and relativistic effects manifest themselves via exchange interaction with the $s$ and $p_{1/2}$ orbitals of the atomic core. Therefore, the $7p_{3/2}$, $6d_{3/2}$ and $6d_{5/2}$ orbitals of Fm are farther from the nucleus than the corresponding orbitals of Er. The trend is further illustrated by the data in Table~\ref{t:spd}, which presents single-electron HF energies of the considered states, including also the $4f$ and $5f$ states. Note that both Fig.~\ref{f:ErFm} and Table~\ref{t:spd} show that the difference between Er and Fm is relatively small. This means that the spectra of the two atoms should be similar and that numerical uncertainty in calculated data for Fm is similar to that of Er.

Table~\ref{t:Fm} shows the calculated odd-parity states of Fm, which are connected to the ground state via electric dipole transitions. The corresponding transition rates are also presented to indicate which transitions might be easier to observe. There is good agreement with available experimental data and MCDF calculations~\cite{Fm1,Fm2}. However, it is clear that a large number of states were missed in previous studies.

Table~\ref{t:Fm} also presents calculated values of the Land\'{e} \textit{g}-factors. They are used to generate the state labels by comparing the calculated \textit{g}-factors to the non-relativistic expression (\ref{e:g}) (see explanation below formula (\ref{e:g})).

Comparison of the data in Tables~\ref{t:Er} and \ref{t:Fm} indicates that the spectra of the two atoms are very similar indeed. There are some differences, which should be attributed to relativistic effects. E.g., the gap between the ground and first excited states is larger in Fm than in Er. Also, the spread of energies within one configuration is larger  in Fm than in Er. The latter is most probably due to larger fine structure of $p$, $d$ and $f$ states.

\begin{table*}
		\caption{\label{t:Fm}
			Excitation energies ($E$, cm$^{-1}$), transition amplitudes ($A$$_{ab}$, a.u.), electric dipole transition rates to the ground state ($T$$_{ab}$, s$^{-1}$), and $g$-factors for some low odd states of Fm atom with \textit{J}= 5,6,7. The difference between the Expt. and present energies is given by $\Delta$ = $\textit E$$_{\rm Expt.}$ - ${\textit E}$$_{\rm present}$. $J^p$ stands for the total angular momentum and parity.} 
		\begin{ruledtabular}
			\begin{tabular}{ccccccccccccc}
				&&&
				\multicolumn{5}{c}{$E$}&
				\multicolumn{1}{c}{$A$$_{ab}$}&
				\multicolumn{3}{c}{$T$$_{ab}$}&
				\multicolumn{1}{c}{\textit{g}}\\
				\cline{4-8}
				\cline{9-9}
				\cline{10-12}
				\cline{13-13}
				\multicolumn{1}{c}{\textit{N}}& 
				\multicolumn{1}{c}{Conf.}&
				\multicolumn{1}{c}{$J^p$}&
				\multicolumn{1}{c}{Present}&
				\multicolumn{2}{c}{MCDF\tablenotemark[1]}&
				\multicolumn{1}{c}{Expt.\tablenotemark[1]}&
				\multicolumn{1}{c}{$\Delta$}&
				\multicolumn{1}{c}{Present}&
				\multicolumn{1}{c}{Present}&
				\multicolumn{1}{c}{MCDF\tablenotemark[1]}&
				\multicolumn{1}{c}{Expt.\tablenotemark[1]}&
				\multicolumn{1}{c}{Present} \\
				\hline
				1 & $5f^{12}7s^2$     & $6^+$ & 0 & $^3$H$_6$ & 0&0&0&0&0&0&0& 1.1619 \\
				2 & $5f^{11}7s^26d$   & $6^-$ & 18915 &&$>$13000\tablenotemark[2]&&&-0.3124&1.0283$\cdot$10$^5$& && 1.2880  \\
				3 & $5f^{12}7s7p$     & $6^-$ & 19012 &&&&&1.3588&1.9756$\cdot$10$^6$& && 1.2594  \\
				4 & $5f^{11}7s^26d$   & $7^-$ & 19808 &&&&&0.3820&1.5304$\cdot$10$^5$& && 1.2378  \\
				5 & $5f^{12}7s7p$     & $6^-$ & 20077  &&&&&3.0391&1.1638$\cdot$10$^7$& && 1.0843  \\
				6 & $5f^{12}7s7p$     & $7^-$ & 20399  &&&&&3.8242&1.6751$\cdot$10$^7$& && 1.1921  \\
				7 & $5f^{12}7s7p$     & $5^-$ & 20711   &&&&&-3.0248&1.4957$\cdot$10$^7$& && 1.1527  \\
				8 & $5f^{11}7s^26d$  & $5^-$ & 23175 &&&&&0.1828&7.6499$\cdot$10$^4$& && 1.1747  \\
				9 & $5f^{11}7s^26d$   & $6^-$ & 23761 &&&&&-1.1413&2.7206$\cdot$10$^6$& && 1.1635  \\
				10 & $5f^{11}7s^26d$   & $7^-$ & 23858 &&&&&-1.0693&2.0954$\cdot$10$^6$& && 1.1527  \\
				11 & $5f^{12}7s7p$   & $5^-$ & 24294 &&&&&-0.9523&2.3928$\cdot$10$^6$& && 1.2003  \\
				12 & $5f^{12}7s7p$     & $7^-$ & 25202 &&&&&0.0250&1.3486$\cdot$10$^3$& && 1.2351  \\
				13 & $5f^{12}7s7p$     & $6^-$ & 25442 &$^5$I$^o_6$ &25226&25099.8$\pm$0.2 &-342&-1.3421&4.6186$\cdot$10$^6$&1.89$\cdot$10$^6$ &(3.4$\pm$0.8)$\cdot$10$^6$ & 1.1861\\
				14 & $5f^{12}7s7p$     & $5^-$ & 25494 &$^5$G$^o_5$ &25471&25111.8$\pm$0.2 &-382&-0.9201&2.5808$\cdot$10$^6$&1.28$\cdot$10$^6$ &(3.5$\pm$0.7)$\cdot$10$^6$ & 1.1204  \\
				15 & $5f^{12}7s7p$     & $5^-$ & 28520 &$^3$G$^o_5$ &27633&27389$\pm$1.5 &-1131&-5.1793&1.1451$\cdot$10$^8$&1.98$\cdot$10$^8$ && 1.2120  \\
				16 & $5f^{12}7s7p$     & $6^-$ & 28662 &$^3$H$^o_6$ &27394&27466$\pm$1.5 &-1196&3.0233&3.3511$\cdot$10$^7$&2.43$\cdot$10$^8$ &$\geq$2.9$\cdot$10$^6$ & 1.2485  \\
				17 & $5f^{12}7s7p$     & $5^-$ & 28690 &&28540&28185$\pm$1.5 &-505&6.0863&1.6097$\cdot$10$^8$&2.82$\cdot$10$^5$ && 1.2029  \\
				18 & $5f^{12}7s7p$     & $6^-$ & 28995 &&&28377$\pm$1.5 &-618&7.8270&2.3251$\cdot$10$^8$& && 1.1744  \\
				19 & $5f^{12}7s7p$     & $7^-$ & 29348 &$^3$I$^o_7$ &27802&28391$\pm$1.5 &-957&9.7202&3.2227$\cdot$10$^8$&3.67$\cdot$10$^8$ &$\geq$1.1$\cdot$10$^7$ & 1.1455  \\
				20 & $5f^{11}7s^26d$   & $5^-$ & 30236 &&29359&&&-0.0602&1.8429$\cdot$10$^4$&3.58$\cdot$10$^7$ && 1.1338  \\
				21 & $5f^{11}7s^26d$   & $6^-$ & 30943 &&&&&-0.9144&3.8572$\cdot$10$^6$& && 1.0814  \\
				22 & $5f^{11}7s^26d$   & $5^-$ & 31953 &&&&&0.2406&3.4736$\cdot$10$^5$& && 1.1369  \\
				23 & $5f^{12}7s7p$   & $5^-$ & 32200 &&&&&0.4368&1.1719$\cdot$10$^6$& && 1.1124  \\
				24 & $5f^{12}7s7p$   & $5^-$ & 32239 &&&&&-0.1187&8.6814$\cdot$10$^4$& && 1.0473  \\
				25 & $5f^{11}7s^26d$   & $6^-$ & 32583 &&&&&0.4648&1.1635$\cdot$10$^6$& && 1.1403  \\
				26 & $5f^{11}7s^26d$   & $7^-$ & 32693 &&&&&-0.9598&4.3436$\cdot$10$^6$& && 1.0969  \\
				27 & $5f^{12}7s7p$   & $5^-$ & 33093 &&&&&0.0298&5.9416$\cdot$10$^3$& && 1.0494  \\
				28 & $5f^{11}7s^26d$   & $7^-$ & 33328 &&&&&0.7106&2.5223$\cdot$10$^6$& && 1.0942  \\
				29 & $5f^{11}7s^26d$   & $5^-$ & 33575 &&&&&0.0410&1.1680$\cdot$10$^4$& && 1.1448  \\
				30 & $5f^{12}7s7p$   & $6^-$ & 33656 &&&&&0.3889&8.9755$\cdot$10$^5$& && 1.0886  \\
				31 & $5f^{11}7s^26d$   & $6^-$ & 33750 &&&&&0.8209&4.0337$\cdot$10$^6$& && 1.0760  \\
				32 & $5f^{11}7s^26d$   & $6^-$ & 35020 &&&&&-1.5562&1.6194$\cdot$10$^7$& && 1.0968  \\
				33 & $5f^{11}7s^26d$   & $7^-$ & 35106 &&&&&1.7151&1.7174$\cdot$10$^7$& && 1.0834  \\
				34 & $5f^{11}7s^26d$   & $5^-$ & 35442 &&&&&0.8021&5.2698$\cdot$10$^6$& && 1.0987  \\
				35 & $5f^{11}7s^26d$   & $6^-$ & 36587 &&&&&2.4329&4.5135$\cdot$10$^7$& && 1.1463  \\
\end{tabular}
%			\begin{minipage}{6.5 cm}%
%				\tiny  $^{c,d}$ Ref. [7,8]; $^{e}$Ref. [9].%
%			\end{minipage}%
\end{ruledtabular}
\tablenotetext[1]{Refs.~\cite{Fm1,Fm2}}

\tablenotetext[2]{Ref.~\cite{Sato}}
	\end{table*}

\begin{table}
\caption{\label{t:spd} Single-electron energies (in a.u.) of the lowest valence orbitals of Er and Fm.}
\begin{ruledtabular}
\begin{tabular}{llll}	
\multicolumn{2}{c}{Er}&
\multicolumn{2}{c}{Fm}\\
\multicolumn{1}{c}{Orbital}&
\multicolumn{1}{c}{Energy}&
\multicolumn{1}{c}{Orbital}&
\multicolumn{1}{c}{Energy}\\
\hline
$4f_{5/2}$  & -0.7576 & $5f_{5/2}$ & -0.7509 \\
$4f_{7/2}$  & -0.7121 & $5f_{7/2}$ & -0.6711 \\
$6s_{1/2}$  & -0.4065 & $7s_{1/2}$ & -0.4269 \\
$6p_{1/2}$  & -0.1210 & $7p_{1/2}$ & -0.1240 \\
$6p_{3/2}$  & -0.1140 & $7p_{3/2}$ & -0.1057 \\
$5d_{3/2}$  & -0.0837 & $6d_{3/2}$ & -0.0819 \\
$5d_{5/2}$  & -0.0832 & $6d_{5/2}$ & -0.0809 \\
\end{tabular}
\end{ruledtabular}
\end{table}

\subsection{Ionization potential}

The first ionization potential (IP) of an atom is calculated as a difference in the ground state energy between the neutral atom and the singly-positive ion. The calculations are the same as those for transition energies. Our value for the IP of Er is 6.102~eV, which is in excellent agreement with the experimental value 6.1077~eV~\cite{NIST}. Similarly, the calculated IP of Fm, 6.559~eV, is in very good agreement with the experimental value, 6.52(13)~eV~\cite{Sato}. In contrast to the calculation of transition energies, where very little published data can be found, calculation of the IP of Fm has been performed by many authors. A detailed review of the results can be found in Ref.~\cite{Sato}. Our result is the closest to the experimental value. Another very accurate result has been obtained with the CCSD(T) method in Ref.~\cite{Sato}. Its value is 6.469~eV. Our value is only about 1\% larger.
The data are summarised in Table~\ref{t:IP}.
	\begin{table*}
		\caption{\label{t:IP}
			Experimental and theoretical values of the first ionization potential IP$_1$ (eV).} 
		\begin{ruledtabular}
			\begin{tabular}{ccccccccc}
				&
				\multicolumn{2}{c}{State}&
				\multicolumn{5}{c}{IP$_1$}\\
				\cline{2-3}
				\cline{4-9}
				
				\multicolumn{1}{c}{Atoms}&
				\multicolumn{1}{c}{Initial}&
				\multicolumn{1}{c}{Final}&
				\multicolumn{1}{c}{Presesnt}&
				\multicolumn{1}{c}{CCSD(T)} \tablenotemark[1] &
				\multicolumn{1}{c}{MCDF} \tablenotemark[1] &
				\multicolumn{1}{c}{QR PP+ACPF} \tablenotemark[2] &
				\multicolumn{1}{c}{Semiempirical} \tablenotemark[3]&
				\multicolumn{1}{c}{Expt} \\
				
				\hline
				Er & $4f^{12}6s^2 \ ^6$H$_6$ & $4f^{12}6s \ (6,1/2)_{13/2}$&6.1017&&5.94&5.89&&6.1077 \tablenotemark[4] \\
				Fm & $5f^{12}7s^2 \ ^6$H$_6$ & $5f^{12}7s \ (6,1/2)_{13/2}$&6.5587&6.469&6.22&6.26&6.50&6.52$\pm$0.13 \tablenotemark[1] \\

			\end{tabular}

		\end{ruledtabular}
	\tablenotetext[1]{Ref.~\cite{Sato}}
		\tablenotetext[2]{Ref.~\cite{Liu1,Liu2}}
		\tablenotetext[3]{Ref.~\cite{J. Sugar}}
	\tablenotetext[4]{Ref.~\cite{NIST}}
	
	\end{table*}

\subsection{Scalar polarizabilities of Er and Fm atoms}

Scalar polarizabilities of Er and Fm were calculated in Ref.~\cite{pol}. In this work we have an opportunity to calculate the polarizability in a completely different way and to compare the results.
Scalar polaizabilities $\alpha_0$ are given by (we use atomic units)
\begin{equation}\label{e:pol}
	\alpha_0=\dfrac{2}{3(2J_a+1)}\sum_{n}\frac{A_{an}^2}{E_{a}-E_{n}},
\end{equation}
where $J_a$ is the total angular momentum of the ground state of the atom ($J_a$=6 for Er and Fm),  $A_{an}$ are amplitudes (reduced matrix elements) of the electric dipole transitions from the ground state to odd excited states; $E_a-E_n$ is an excitation energy.
Summation goes over the complete set of odd states. We use the amplitudes and energies from Tables~\ref{t:Er} and \ref{t:Fm} to perform the calculations. The results are presented in Table~\ref{t:pol}. 

There are some significant differences in approaches used to calculate polarizabilities in the present work and in Ref.~\cite{pol}. In Ref.~\cite{pol} the $4f$ and $5f$ electrons were attributed to the core and the atoms were treated as two-valence electrons systems with the $6s^2$ ground state configuration for Er and $7s^2$ ground state configuration for Fm. This means that only $s-p$ electric dipole matrix elements were used in the calculations of the polarizabilities. On the other side, the summation was extended to very high energy states to ensure the saturation of the summation. In the present work we use all possible kinds of transitions, but the summation is truncated at much lower number of levels. It is limited to states presented in Tables~\ref{t:Er} and \ref{t:Fm}. 

Note that the total polarizability is the sum of the core and valence contributions. In the present work we do not calculate the core contributions but take them from Ref~\cite{pol}. There is some difference in the definitions of the core in these two works. We treat the $4f$ and $5f$ electrons as being in the valence space while in \cite{pol} they were attributed to the core. This should lead to some differences in the core polarizabilities. However, the contribution of $f$-electrons to the polarizabilities 
is small and can be neglected \cite{pol}. Another source of uncertainty is the use of calculated energies in (\ref{e:pol}). As can be seen from Table~\ref{t:Er} the difference between theoretical and experimental energies is significant for some states. This may lead to significant errors in the polarizability. 
To check this we performed calculations twice, using theoretical energies in one run and experimental energies in the second run. Both results are presented in Table~\ref{t:pol}. The difference between them is just 0.5\%. The difference between the total polarizabilities of present work and earlier calculations of Ref.~\cite{pol} is 1\% for Er and 10\% for Fm. 

Another way of estimation uncertainty is by replacing calculated transition amplitudes by experimental values when they are available. 
Note, however, that accuracy in a calculation of polarizability is usually much better than accuracy of calculation of transition rates, especially for weak transitions.
Indeed, transition rates are very sensitive to mixing between close excited states (which is usually very strong) while polarizability is not affected by mixing of (nearly) degenerate excited states. We just need a complete set of basis states. For the same reason, replacing part of transition amplitudes by experimental values is not a way to improve accuracy (see, e.g.~\cite{Yb-pol} for a more detailed discussion). In the end, we found that the uncertainty for Er is about 10\%, while for Fm it might be larger.
		
\begin{table}
\caption{\label{t:pol}
Scalar polarizabilities of Er and Fm atoms (in $a_B^3$).} 
\begin{ruledtabular}
\begin{tabular}{lcddddd}
				&&
				\multicolumn{3}{c}{Ref.~\cite{pol}}&
				\multicolumn{2}{c}{Present Work }\\
				
				\cline{3-5}
				\cline{6-7}
				
				\multicolumn{1}{c}{Atom}&
				\multicolumn{1}{c}{Conf.}&

\multicolumn{1}{l}{Core}&
\multicolumn{1}{c}{Valence}&
\multicolumn{1}{c}{Total}&
\multicolumn{1}{c}{Valence }&				
\multicolumn{1}{c}{Total}\\
\hline

				Er &$4f^{12}6s^{2}$&6.3&143.9&150.2&145.66&151.96\\
				Er&$4f^{12}6s^{2}$&6.3&&&144.94\tablenotemark[1]&151.24\\
				Fm&$5f^{12}7s^{2}$& 8.4 &105&113.4&115.92&124.32\\
				
\end{tabular}			
\end{ruledtabular}
\tablenotetext[1]{This value was obtained when NIST energies were used.}

\end{table}

\subsection{Hyperfine structure}

Studying hyperfine structure is a way to get information about the nucleus by extracting nuclear moments from the comparison between theory and measurements. There are many accurate published data on hfs of  $^{167}$Er~\cite{Cheng85,Childs83,Childs86}.
There are also measurements of the hfs for $^{255}$Fm~\cite{Fm2}. Therefore, we perform the calculations for both atoms using erbium as a testing ground to check the accuracy of the method. We calculate magnetic dipole hfs constant $A$ and electric quadrupole hfs constant $B$ using the RPA and CIPT methods (see Eqs.~(\ref{e:RPA}) and (\ref{e:A})). The results for $^{167}$Er are presented in Table~\ref{t:Erhfs}. One can see that we have an excellent agreement between theory and experiment for magnetic dipole constant $A$ for all states of the ground configuration. The difference does not exceed 3\%, and for some states it is even smaller. For electric quadrupole constant $B$ the difference between theory and experiment is within 10\% for most states except the second one. Here the discrepancy, by a factor of two, is probably due to anomalously small value of the constant, which might be a result of cancellation between different contributions. However, the absolute difference between theory and experiment ($\sim$ 500~MHz) is about the same as for the ground state.
The accuracy is almost equally good for the states of the $4f^{12}6s6p$ configuration (marked with letter P in Table~\ref{t:Erhfs}), while it is somewhat worse for the states of the $4f^{11}6s^25d$ configuration (marked with letter D in Table~\ref{t:Erhfs}). In the end, it is clear that the accuracy is sufficiently good to use the calculations for extracting nuclear moments.

\begin{table*}
\caption{\label{t:Erhfs}
Hyperfine structure constants $A$ and $B$ (in MHz) of $^{167}$Er. Nuclear spin $I=7/2$, nuclear magnetic moment $\mu(^{167}{\rm Er})=-0.56385(12)\mu_N$~\cite{Stone1};
nuclear electric quadrupole moment $Q(^{167}{\rm Er})=3.57(3)~b$~\cite{Stone2}; $g_I=\mu/I$. Letters S, D, P indicate the leading configurations, $4f^{12}6s^2$, $4f^{11}6s^25d$, and $4f^{12}6s6p$, respectively. The last column gives references to experimental data.} 
\begin{ruledtabular}
\begin{tabular}{l dddddddd l}
\multicolumn{1}{c}{Conf.}&
\multicolumn{2}{c}{Energy (cm$^{-1}$)}&
\multicolumn{2}{c}{$A_{\rm theor}$}&
\multicolumn{1}{c}{$A_{\rm expt}$}&
\multicolumn{2}{c}{$B_{\rm theor}$}&
\multicolumn{1}{c}{$B_{\rm expt}$}&
\multicolumn{1}{c}{Ref.}\\
\cline{2-3}
&\multicolumn{1}{c}{NIST\cite{NIST}}&
\multicolumn{1}{c}{CIPT}\\

\hline
%N Type                 A(gi)  A     A_Expo    B(Q)   B     B_expt
   S &    0.000     &    0  & 724g_I & -117  & -120.487  & -1413Q & -5037  & -4552.984 & \cite{Cheng85} \\
   S &  5035.193  & 5364 &  752g_I  & -122 & -121.9    &  295Q & 1050  &  516    & \cite{Childs83} \\
   S  & 6958.329  & 7275 &  976g_I   & -158  & -159.4   & -1273Q & -4539 & -4120  &    \cite{Childs83} \\
   S  &10750.982 & 10875 & 1075g_I  & -174  & -173.4   &  -729Q & -2600 & -2429   &   \cite{Childs83} \\
   S  &12377.534 & 13353  & 827g_I   & -139  & -143.4    &  373Q & 1331  & 1236   &   \cite{Childs83} \\
   S  &13097.906 & 14624 & 1063g_I   & -172  & -167.2    &  460Q  & 1640  & 1688   &   \cite{Childs83} \\

   D   &7176.503  & 5419  & 833g_I   & -135  & -139.957  & -464Q & -1655  & -709.396 &  \cite{Childs86} \\
   D   &7696.956  & 5959  & 705g_I   & -114  & -125.851  & -626Q & -2230  & -3046.052 & \cite{Childs86} \\
   D   &9350.106  & 7584  & 643g_I   & -104  & -119.870  & -665Q & -2372  & -3062.704 & \cite{Childs86} \\
   D   &8620.565  & 4771  & 613g_I   & -99  & -113.582  & -486Q & -1733  & -782.987 & \cite{Childs86} \\
  
   P  &17157.307 & 17342 & 1071g_I  & -173 & -172.5   & -1400Q & -4992 & -4440   &  \cite{Childs83} \\
\end{tabular}			
\end{ruledtabular}
\end{table*}

The results of similar calculations for fermium are presented in Table~\ref{t:Fmhfs}. It is natural to expect that due to similar electronic structure the accuracy for fermium is similar to what we have for erbium. As one can see, the values of the hfs constants for the ground state are also similar. For example, the magnetic dipole constant $A$ for the ground state of Fm (655$g_I$~MHz) is only 14\% smaller than for Er (724$g_I$~MHz). The smaller value can be explained in the following way. Only $f$-electrons contribute to the hfs of the ground state. This is because $s$-electrons form a closed subshell ($6s^2$ for Er and $7s^2$ for Fm) with zero total angular momentum. The $4f$-electrons of Er are deeper in the core on the energy scale than the $5f$-electrons of Fm (see Table~\ref{t:spd}). Therefore, the overlap of the $4f$ function with the $s$ functions of the core is larger than for the $5f$ state. This means that the RPA correction to the hfs is also larger for Er than for Fm.
%the $4f$ density in the vicinity of the nucleus is larger than the density of $5f$-electrons, and the corresponding hfs constant $A$ is larger, too. 
Similarly, the electric quadrupole constant $B$ for Fm ($-1750Q$~MHz) is only 24\% larger than the constant $B$ for Er ($-1413Q$~MHz) due to the larger gradient of electric field. There are also similarities between the hfs constants of the $4f^{12}6s6p$ and $5f^{12}7s7p$ configurations of Er and Fm. However, they are less pronounced due to the fact that the corresponding states of Fm are significantly higher on the energy scale than in Er. In the end, we see no reason to believe that the accuracy of the calculations for Fm is lower than for Er.

The extraction of nuclear moments from the comparison of theoretical and experimental hfs constants presents a problem. There is no way to get consistent results from the three states (see Table~\ref{t:Fmhfs}). The most likely reason in our view may come from the inaccurate interpretation of measured hfs in terms of six hfs constants, the $A$ and $B$ constants for the three states. Ref.~\cite{Fm1} presents two quite different interpretations, which are both included in Table~\ref{t:Fmhfs}, and states that more interpretations are possible under different assumptions.  We hope that our calculations would help to re-evaluate the experimental data.

\begin{table*}
\caption{\label{t:Fmhfs}
Hyperfine structure constants $A$ and $B$ (in MHz) of $^{255}$Fm. Two different interpretations of the experimental data are taken from Ref.~\cite{Fm1}.
Notations R1 and R2 are also taken from Ref.~\cite{Fm1}.}
\begin{ruledtabular}
\begin{tabular}{l dr c dd c dd}
\multicolumn{1}{c}{Conf.}&
\multicolumn{2}{c}{Energy (cm$^{-1}$)}&
\multicolumn{1}{c}{$A_{\rm theor}$}&
\multicolumn{2}{c}{$A_{\rm expt}$}&
\multicolumn{1}{c}{$B_{\rm theor}$}&
\multicolumn{2}{c}{$B_{\rm expt}$}\\
\cline{2-3}
&\multicolumn{1}{c}{Expt\cite{Fm2}}&
\multicolumn{1}{c}{CIPT}\\
\hline

GS &  0.0           &   0    &  655$g_I$ & -320 & -760 & -1750$Q$ & -22000 & -8700 \\
R1  & 25099.80 & 25442 & 1810$g_I$ &-530 & -660 & -1975$Q$ & -2900 & -14800 \\
R2  & 25111.80 & 25494  &  195$g_I$  & -690 & -730 & -1823$Q$ & -1700 & -390 \\
\end{tabular}			
\end{ruledtabular}
\end{table*}

\subsection{Isotope shift}

Only the field shift needs to be considered in heavy elements. The mass shift is small and can be neglected~\cite{No-IS,No-nuc}.
To calculate the IS for Fm we use nuclear parameters that come from nuclear calculations~\cite{Afanasjev}. We use only two most important parameters, the nuclear RMS radius $r_p$ and the parameter of nuclear quadrupole deformation $\beta$. The nuclear charge density is approximated by the formula
\begin{equation}
\rho(r,\theta) = \frac{a}{1+ \exp{\frac{r-r_n(\theta)}{t}}}.
\label{e:Fermit}
\end{equation}
Here $a$ is a normalization constant, $t$ is the nuclear skin thickness (we use the standard value $t= 2.3/4\ln(3)$ {\rm fm}), $r_n(\theta)$ is a variable nuclear radius, $r_n(\theta) = r_0(1+\beta Y_{20}(\theta)$), where \textit{Y}$_{20}$ is a spherical harmonic.
Electrons feel the nuclear density averaged over nuclear rotations, $\rho(r) = \int_0^{\pi} \rho(r,\theta)d\theta$. The constant $a$ in (\ref{e:Fermit}) is chosen to get $\int \rho dV = Z$. The parameter $r_0$ in the variable nuclear radius is treated as a fitting parameter to obtain the correct value of the nuclear RMS radius. The resulting averaged nuclear charge density $\rho(r)$ is used to calculate the nuclear potential. The IS is calculated as a difference between transition frequencies of two isotopes. 
It was suggested in Refs.~\cite{E120,No-nuc} that the IS for a wide range of values for nuclear RMS radii and quadrupole deformation $\beta$ can be approximated by a formula containing powers of $\delta \langle r^2 \rangle$ and $\Delta \beta$. It was also suggested that for neighbouring isotopes one can keep only the leading terms
\begin{equation}\label{e:Fd}
%\delta \nu = F\delta\langle r^2\rangle +b \Delta \beta + c \Delta \beta \delta\langle r^2\rangle +d \Delta \beta^2 + e (\delta\langle r^2\rangle)^2.
\delta \nu = F\delta\langle r^2\rangle  +d \Delta \beta.
\end{equation}
%Here $F,b,c,d,e$ are fitting parameters chosen to fit the calculated values of IS for several isotopes. 
Here the first term is just the standard formula for field IS. Note that relativistic consideration leads to a different formula for field isotope sift, 
$\delta \nu = F^{\prime}\delta\langle r^{2\gamma}\rangle$, where $\gamma = \sqrt{1-(\alpha Z)^2}$~\cite{gamma1,gamma2}. 
The relativistic formula can be presented as an expansion $F^{\prime}\delta\langle r^{2\gamma}\rangle = F\delta\langle r^2\rangle + G\delta\langle r^2\rangle^2 + \dots$. For neighbouring isotopes higher-order corrections are small and can be neglected, brining us back to (\ref{e:Fd}). 
The second term in (\ref{e:Fd}) takes into account nuclear deformation. 
The advantage of using a non-relativistic formula is the ability to extract the value of $\delta\langle r^2\rangle$ from the isotope shift measurements. 
The disadvantage is the loss of the universality. Constants $F$ and $d$ in (\ref{e:Fd}) are isotope-dependent and should be calculated from scratch for a new pair of isotopes if the number of neutrons is significantly different.
 The values of $F$ and $d$ are found from fitting the calculated IS. 
% Note that $F$ and $d$ in (\ref{e:Fd}) are isotope-dependent, and for a pair of isotopes with significantly different neutron numbers, the evaluation of $F$ and $d$ should be done from scratch. 
 If the IS is measured for at least two atomic transitions, then (\ref{e:Fd}) can be used to determine both $\delta\langle r^2\rangle$  and $\Delta \beta$. Our calculated values of $F$ and $d$ for IS around the $^{255}$Fm isotope are presented in Table~\ref{t:Fd}.
Note that, according to the nuclear theory~\cite{Afanasjev}, nuclei in this region are deformed with $\beta \sim 0.3$, and $\Delta \beta \sim 0.02$ for neighbouring isotopes. Using the data from Table~\ref{t:Fd}, we estimate that the contribution of nuclear deformation to the IS is $\sim 0.01~{\rm cm}^{-1}$. This value is large enough to be detected. It means that measurements of IS can be used to study nuclear deformations. For odd isotopes like $^{255}$Fm the use of IS to study nuclear deformation is complementary to the measurements of the electric quadrupole hfs. However, for even-even isotopes, such as $^{254}$Fm, $^{256}$Fm, which have no hfs, this is a unique way of studying nuclear deformation by means of atomic spectroscopy.

\begin{table}
\caption{\label{t:Fd}Parameters of formula (\ref{e:Fd}) for isotope shifts for transitions from the ground state to excited odd states of $^{255}$Fm.
$N$ is state number from Table~\ref{t:Fm}, letters P and D indicate dominating configurations, $5f^{12}7s7p$ and $5f^{11}7s^26d$ respectively. }
\begin{ruledtabular}
\begin{tabular}{r cc cdd}
\multicolumn{1}{c}{$N$}&
\multicolumn{1}{c}{Conf.}&
\multicolumn{1}{c}{$J$}&
\multicolumn{1}{c}{$E$}&
\multicolumn{1}{c}{$F$}&
\multicolumn{1}{c}{$d$}\\
&&&
\multicolumn{1}{c}{cm$^{-1}$}&
\multicolumn{1}{c}{cm$^{-1}$/fm$^2$}&
\multicolumn{1}{c}{cm$^{-1}$}\\
\hline
 7 & P  & 5 &   20711 &  -2.92 &    0.423 \\
 8 & D  & 5 &   23175 &   0.78 &    0.610 \\
11 & P  & 5 &   24294 &  -0.19 &    0.585 \\
14 & P  & 5 &   25494 &  -3.14 &    0.397 \\
15 & P  & 5 &   28520 &  -2.53 &    0.294 \\
17 & P  & 5 &   28690 &  -1.24 &    0.331 \\
     	       	             
 2 & P  & 6 &   18915 &  -2.95 &    0.455 \\
 3 & D  & 6 &   19012 &   3.18 &    0.604 \\
 5 & P  & 6 &   20077 &  -2.88 &    0.443 \\
 9 & D  & 6 &   23761 &   2.95 &    0.741 \\
13 & P  & 6 &   25442 &  -3.21 &    0.256 \\
16 & P  & 6 &   28662 &  -3.11 &    0.404 \\
18 & P  & 6 &   28995 &  -1.29 &    0.230 \\
     	       	             
 4 & D  & 7 &   19808 &   3.05 &    0.602 \\
 6 & P  & 7 &   20399 &  -2.59 &    0.431 \\
10 & D  & 7 &   23858 &   3.31 &    0.745 \\
12 & P  & 7 &   25202 &  -3.14 &    0.399 \\
19 & P  & 7 &   29348 &  -2.04 &    0.202 \\
\end{tabular}
\end{ruledtabular}
\end{table}

Finally, in Table~\ref{t:FmIS} we present the calculated IS between previously studied isotope $^{255}$Fm~\cite{Fm1,Fm2} and an isotope $^{284}$Fm which has a "magic" number of neutrons ($N=184$) and thus a spherically-symmetric nucleus~\cite{Afanasjev}. We keep non-relativistic labelling of the states used in~\cite{Fm1,Fm2} for easy comparison.
The shift was calculated as a difference between separate calculations of transition frequencies for $^{255}$Fm and  $^{284}$Fm. 
Nuclear RMS radii  ($r_p$) were taken from nuclear calculations, $r_p$=5.976~fm for $^{255}$Fm and $r_p$=6.063~fm for $^{284}$Fm~\cite{Afanasjev}.
The $^{255}$Fm isotope is used as a reference.
% Formula (\ref{e:Fd}) was not used.
These data can be useful in the search for metastable isotopes, e.g. in the astrophysical data~\cite{ISinStars}. 
		
	\begin{table}
		\caption{\label{t:FmIS}
			Isotope shift (IS, cm$^{-1}$) between the experimentally studied isotope $^{255}$Fm and the isotope with the magic number of neutrons,  $^{284}$Fm. Calculations for $^{255}$Fm were done with nuclear RMS radius $r_p$=5.976~fm, while for $^{284}$Fm $r_p$=6.063~fm~\cite{Afanasjev}.
			The last column is the difference between excitation energies for  $^{284}$Fm and  $^{255}$Fm.}
		
		\begin{ruledtabular}
			\begin{tabular}{cccccc}
				&&&
				\multicolumn{2}{c}{Energy (cm$^{-1}$)}\\

				\cline{4-5}
				
				\multicolumn{1}{c}{N}& 
				\multicolumn{1}{c}{Conf.}&
				\multicolumn{1}{c}{\textit{LSJ}}&
				\multicolumn{1}{c}{$^{255}$Fm}&
				\multicolumn{1}{c}{$^{284}$Fm}&
				\multicolumn{1}{c}{IS} \\
				\hline
				
				13 & $5f^{12}7s7p$     & $^{5}$I$^o_6$ & 25442 &25435& -7\\
				14 & $5f^{12}7s7p$     & $^5$G$^o_5$ & 25494 &25487& -7 \\
				15 & $5f^{12}7s7p$     & $^3$G$^o_5$ & 28520 &28515& -5 \\
				16 & $5f^{12}7s7p$     & $^3$H$^o_6$ & 28662 &28656&-6  \\
				17 & $5f^{12}7s7p$     & $^3$G$^o_5$ & 28690 &28685&-5\\
				19 & $5f^{12}7s7p$     & $^3$I$^o_7$ & 29348 &29343& -5 \\
				20 & $5f^{11}7s^26d$   & $^9$I$^o_5$ & 30236 &30243&~7  \\
				
			\end{tabular}
			
		\end{ruledtabular}
	\end{table}

\section{Conclusions}

All odd energy levels of Fm within the optical range ($E < 40000~{\rm cm}^{-1}$) are calculated with an uncertainty of a few hundred cm$^{-1}$. The results are in good agreement with previous theoretical and experimental studies where the data are available. Twenty eight new levels are reported. Transition rates, hyperfine structure, static scalar polarizabilities, ionization potentials, and isotope shifts are also calculated. Transition rates might be useful for planning further experimental studies. Calculation of the hfs indicates the need to re-visit the interpretation of experimental data for Fm. Isotope shift data can be used for extracting information on the nuclear deformation and change of the nuclear RMS radius from the IS measurements. They can also be used in the search for metastable Fm isotopes with spherically-symmetric nuclei. 

%\section{Acknowledgements}
\acknowledgements
The authors are grateful to Hartmut Backe for stimulating discussions.
	This work was supported by the Australian Research Council.
	JGL acknowleged the financial support from the National Natural Science Foundation of China (Grant No. 11874090).
	VAD would like to express special thanks to the Institute of Applied Physics and Computational Mathematics in Beijing for its hospitality and support. 
	This research includes computations using the computational cluster Katana supported by Research Technology Services at UNSW Sydney.

\end{document}